\documentclass[american,aps,pra,reprint,superscriptaddress,10pt]{revtex4-1}
\usepackage[T1]{fontenc}
\usepackage[sc]{mathpazo}
\usepackage{amsmath}
\usepackage{amssymb}
\usepackage{enumerate}
\usepackage{amsthm}% theorems and proofs
\usepackage{color}

\newtheorem{lem}{Lemma}

\usepackage{amsfonts,mathrsfs}
\usepackage[style]{fncychap}
\usepackage{graphicx} % For including graphics N.B. pdftex graphics driver
\usepackage{geometry} % allows easy specifying of the page layout
\usepackage{eepic}
\usepackage{ifthen}
\newboolean{ElectronicVersion}
\setboolean{ElectronicVersion}{true} % CHANGE THIS AS REQUIRED

\usepackage[T1]{fontenc}
\usepackage[sc]{mathpazo}
\usepackage{mathdots}
\usepackage{amsmath}
\usepackage{amssymb}
\usepackage{enumerate}
\usepackage{amsthm}% theorems and proofs
\usepackage{amsfonts,mathrsfs}
\usepackage[style]{fncychap}
\usepackage{graphicx} % For including graphics N.B. pdftex graphics driver
\usepackage{geometry} % allows easy specifying of the page layout
\usepackage{eepic}
\usepackage{ifthen}
\usepackage{cases}

\usepackage{amsfonts,mathrsfs}
\usepackage[style]{fncychap}
\usepackage{graphicx} % For including graphics N.B. pdftex graphics driver
\usepackage{geometry} % allows easy specifying of the page layout
\usepackage{eepic}
\usepackage{ifthen}
\usepackage{cases}

\usepackage{hyperref}
\hypersetup{pdfpagemode=UseNone}

\def\be{\begin{equation}}
\def\ee{\end{equation}}
\def\bea{\begin{eqnarray*}}
\def\eea{\end{eqnarray*}}

\newenvironment{mylist}[1]{\begin{list}{}{
    \setlength{\leftmargin}{#1}
    \setlength{\rightmargin}{0mm}
    \setlength{\labelsep}{2mm}
    \setlength{\labelwidth}{8mm}
    \setlength{\itemsep}{0mm}}}
    {\end{list}}

\newcommand{\bra}[1]{\langle#1|}

\newcommand{\ket}[1]{|#1\rangle}

\newcommand{\braket}[1]{\langle#1\rangle}

\theoremstyle{definition}

\numberwithin{equation}{section}
\def\be{\begin{equation}}
\def\ee{\end{equation}}
\def\bea{\begin{equation*}}
\def\eea{\end{equation*}}
\def\bna{\begin{eqnarray*}}
\def\ena{\end{eqnarray*}}
\def\bn{\begin{eqnarray}}
\def\en{\end{eqnarray}}
\def\bpm{\begin{pmatrix}}
\def\epm{\end{pmatrix}}

\newcounter{questionnumber}

\usepackage{chngcntr}
\counterwithout{equation}{section}

\begin{document}\pagestyle{empty}

%\title{ Superposition, coherence, and $\cal PT$ -symmetry}
\title{Manifestation of Superposition and Coherence in $\cal PT$-symmetry through the $\eta$-inner Product}

 \author{Minyi Huang}
 \email{11335001@zju.edu.cn}
 \affiliation{School of Mathematical Sciences, Zhejiang University, Hangzhou 310027, PR~China}

\author{Ray-Kuang Lee}
 \email{Corresponding author: rklee@ee.nthu.edu.tw}
\affiliation{Department of Physics, National Tsing Hua University, Hsinchu 300, Taiwan}
\affiliation{Institute of Photonics Technologies, National Tsing Hua University, Hsinchu 300, Taiwan}

 \author{Junde Wu}
 \email{Corresponding author: wjd@zju.edu.cn}
\affiliation{School of Mathematical Sciences, Zhejiang University, Hangzhou 310027, PR~China}

\begin{abstract}
Through the $\eta$-inner product, we investigate $\cal PT$-symmetric quantum mechanics from the viewpoint of  superposition and coherence theory.
It is argued that $\cal PT$-symmetric quantum systems are endowed with stationary superposition or superposition-free properties.
 A physical interpretation of $\eta$-inner product in $\mathbb C^2$ is given through the Stokes parameters, showing the difference between broken and unbroken $\cal PT$-symmetric quantum systems.
\end{abstract}

\maketitle

\section{Introduction}
%=============================================================================%
It is known that both coherence and superposition are distinctive features in quantum physics, which make quantum theory depart from classical physics.
In different fields of physics, quantum coherence and superposition play an important role.
The investigation of coherence has a long history and recently attracts increasing interests in the resource theory~\cite{aberg2006quantifying,PhysRevLett.113.140401,levi2014quantitative,PhysRevA.94.052336,PhysRevLett.117.030401,PhysRevA.93.022122,PhysRevLett.116.120404,RevModPhys.89.041003}.
A recent notable progress is the generalization of coherence resource theory to superposition resource theory.
Now, we know that superposition can be converted to entanglement, with the analogues of free states and free operations in coherence theory~\cite{PhysRevLett.116.080402,regula2017converting,PhysRevLett.119.230401}.

Instead of conventional quantum mechanics with Hermitian Hamiltonians, non-Hermitian parity-time ($\cal PT$) symmetry was initially introduced to generalize quantum mechanics, which was first established  by Bender and his colleagues in 1998 \cite{bender1998real}.
Here, $\cal P$ is parity operator and $\cal T$ is time-reversal operator.
 Since then, lots of work have been done to investigate $\cal PT$-symmetric quantum systems. An important  theoretic notion is the  metric operator and the $\eta$-inner product introduced by Mostafazadeh~\cite{mostafazadeh2002Pseudo1,mostafazadeh2002Pseudo2,mostafazadeh2002Pseudo3,mostafazadeh2010pseudo}.
 In addition, although special attention should be paid to local $\mathcal{PT}$-symmetric operations on a quantum composite system~\cite{YiChan}, many useful applications can be found in different branches of physics through the new degree of freedom by $\mathcal{PT}$-symmetry~\cite{el2007theory,makris2008beam,guo2009observation,ruter2010observation,schindler2011experimental,bittner2012p}.

Compared with conventional quantum mechanics, $\cal PT$-symmetry quantum theory has some distinct features.
For example, to discuss the evolution of a $\cal PT$-symmetric system,
we need to choose a preferred basis, which inspires us the scenarios in coherence and superposition theories.
In this note, we will discuss the $\cal PT$-symmetry theory from the perspective of coherence and superposition, revealing some internal connections through the $\eta$-inner product.
As an example, in $\mathbb C^2$,  a physical interpretation of this $\eta$-inner product is given through the analogues in optical polarizations.

\section{Preliminaries}
\subsection{Some basic notions of coherence and superposition}
 We first introduce some notions which will be used.
 In the resource theory of coherence,
 % the notions of incoherent states and operations play a fundamental role.
 given a preferred orthonormal basis $\ket{i}_{i=1}^d$, a state $\rho$ is incoherent if
$\rho=\displaystyle\sum_{i=1}^d p_i\ket{i}\bra{i}$, where $p_i$ is the probability distribution. The set of all the incoherent states is usually denoted by
$\cal I$.
%All the state not in $\cal F$ are considered to have some resource of coherence.

%There are many different definitions of free operations in the theory of coherence. In the following discussions, we mainly use the
%concept of incoherent operation.
A Kraus operator $K_n$ is said to be incoherent if for all $\rho\in \cal I$, $\frac{K_n\rho K_n^\dag}{Tr[K_n\rho K_n^\dag]}\in \cal I$.
An incoherent operation is a completely positive trace preserving (CPTP) map having an incoherent
Kraus decomposition.

Similarly, there are concepts of free states and operations in the theory of superposition \cite{PhysRevLett.119.230401}.
Let $\{\ket{c_i}_{i=1}^d\}$ be a normalized, linearly independent and not necessarily orthogonal basis of the Hilbert space $\mathbb C^d$.
A state $\rho$ is superposition-free if $\rho=\displaystyle\sum_{i=1}^d p_i\ket{c_i}\bra{c_i}$, where $p_i$ is the probability distribution.
The set of all the superposition-free states is usually denote by $\cal F$.

A Kraus operator $K_n$ is said to be superposition-free if for all $\rho\in \cal F$,
$\frac{K_n\rho K_n^\dag}{Tr[K_n\rho K_n^\dag]}\in \cal F$.
 A superposition-free operation
is a CPTP map having a
superposition-free Kraus decomposition.

As was proved in \cite{PhysRevLett.119.230401}, for a set of superposition-free Kraus operators $K_m$ such that $\sum K_m^\dag K_m \leqslant I$, there always exist
superposition-free Kraus operators $F_n$ such that $\sum K_m^\dag K_m +\sum F_n^\dag F_n = I$. Hence the trace-decreasing operations
admitting superposition-free Kraus decompositions are also called superposition-free.

\subsection{Introduction to $\cal PT$-symmetry quantum theory}
%Now we give a self-contained introduction to $\cal PT$-symmetry quantum theory, which only
% concerns finite dimensional spaces $\mathbb C^d$.

$\cal PT$-symmetry quantum theory explores the properties of quantum systems, which are governed by a $\cal PT$-symmetric
Hamiltonian $\cal H$.

A parity operator $\cal P$ is a linear operator such that ${\cal P}^2={\cal I}_d$, where ${\cal I}_d$ is the identity operator on $\mathbb{C}^d$.

A time reversal operator $\cal T$ is an anti-linear operator such that ${\cal
T}^2={\cal I}_d$. Moreover, it is demanded that
${\cal PT}={\cal TP}$.

A linear operator $\cal H$ on $\mathbb{C}^d$ is said to be $\cal
PT$-symmetric if ${\cal H}{\cal PT}={\cal PT}{\cal H}$.

In finite dimensional case, a linear operator corresponds uniquely to a matrix and an anti-linear operator
corresponds to the composition of a matrix and
a complex conjugation \cite{uhlmann16}.
Let $A$ be a matrix. Denote $\overline{A}$ the complex conjugation of $A$ and $A^\dag$ the transpose of $\overline{A}$.
Let $P$, $T$ and $H$  be the matrices of ${\cal P}$, ${\cal T}$ and
$\cal H$, respectively. Then the definition conditions of $\cal P$, $\cal T$, $\cal H$ are
$P^2=T\overline{T}=I$, $PT=T\overline{P}$ and $HPT=PT\overline{H}$.

In conventional quantum mechanics, the Hamiltonians are Hermitian and thus are unitarily similar to a real diagonal matrix.
Analogously, $\cal PT$-symmetric Hamiltonians have a canonical form.

\begin{lem}\cite{huang2017embedding}\label{lem1} A finite dimensional operator $\cal H$ is $\cal PT$-symmetric if and only if
 there exists a matrix $\Psi$ such that $\Psi^{-1}H\Psi=J$,
\begin{equation}
J=\begin{pmatrix}
\begin{smallmatrix}
J_{n_1}(\lambda_1,\overline{\lambda}_1)&&&&&\\
&\ddots&&&\\
&&J_{n_p}(\lambda_p,\overline{\lambda}_p)&&&&\\
&&&J_{n_q}(\lambda_q)&&\\
&&&&\ddots&\\
&&&&&J_{n_r}(\lambda_r)
\end{smallmatrix}
\end{pmatrix}\label{cano2},
\end{equation}
where $J_{n_k}(\lambda_{n_k},\overline{\lambda}_{n_k})=
\bpm \begin{smallmatrix}J_{n_k}(\lambda_{n_k})&0\\0& J_{n_k}(\overline{\lambda_{n_k}})\end{smallmatrix}\epm$,
$J_{n_j}(\lambda_{n_j})$ is the Jordan block, $\lambda_{n_1},
\cdots, \lambda_{n_p}$ are complex (and not real) numbers and $\lambda_{n_q},
\cdots, \lambda_{n_r}$ are real numbers. Moreover,
$PT\overline{\Psi}=\Psi K$,
\begin{equation}
K=\begin{pmatrix}
\begin{smallmatrix}
S_2\otimes I_{n_1}&&&&&\\
&\ddots&&&\\
&&S_2\otimes I_{n_p}&&&&\\
&&&I_{n_q}&&\\
&&&&\ddots&\\
&&&&&I_{n_r}
\end{smallmatrix}
\end{pmatrix}\label{cano1},
\end{equation}
where $S_2=
\bpm \begin{smallmatrix}0&1\\1& 0\end{smallmatrix}\epm$, and $I_{n_j}$ is an identity matrix with the same order as $J_{n_j}(\lambda_{n_j})$.

\end{lem}

In fact, it is an important result in matrix analysis that any matrix, which is similar to a real matrix, has the Jordan form (\ref{cano2}).
On the other hand, it can be verified that a $\cal PT$-symmetric matrix is similar to a real matrix, thus having the Jordan form (\ref{cano2}). As for the canonical form of $PT$ in (\ref{cano1}) and the details of the proof, see \cite{huang2017embedding}.

In equation (\ref{cano2}), if all the blocks $J_{n_k}(\lambda_{n_k},\overline{\lambda}_{n_k})$ vanish and all the blocks
$J_{n_j}(\lambda_{n_j})$ are of order one, then $J$ reduces to a real diagonal matrix.
The Hamiltonians similar to real diagonal matrices are referred to as unbroken $\cal PT$-symmetric. The others are said to be
broken.

In the context of $\cal PT$-symmetry theory, the evolution of a state $\rho$ is given by \[\rho(t)={\cal U}(t)\rho{\cal U}^\dag(t),\] where
${\cal U}(t)=e^{-it\cal H}$.
According to the probability interpretation of inner product, ${\cal U}(t)$ should be inner product preserving.
However, it is not true when $\cal H$ is $\cal PT$-symmetric since ${\cal U}(t)$ is not unitary.

To settle the problem, introduce a Hermitian operator $\eta$ and redefine an $\eta-$inner
product by $\braket{\phi_1,\phi_2}_{\eta}=\braket{\phi_1,\eta\phi_2}$,
where $\phi_1$ and $\phi_2$ are two states.
The evolution ${\cal U}(t)$ preserves the $\eta-$inner
product if and only if ${\cal H}^\dag\eta=\eta \cal H$
\cite{mostafazadeh2002Pseudo1,mostafazadeh2002Pseudo2,mostafazadeh2002Pseudo3,mostafazadeh2010pseudo,deng2012general,mannheim2013pt,horn2012matrix}. An operator $\eta$ satisfying this condition is said to be a metric operator of $\cal H$. Moreover, the following lemma gives the canonical
form of a metric operator.

\begin{lem}\cite{gohberg1983matrices2}\label{lem2}
For each $\cal PT$-symmetric operator $\cal H$, there exists invertible Hermitian matrix $\eta$ such that
$H^\dag \eta=\eta H$. Moreover, there exists a matrix $\Psi$ such that $\Psi^{-1}H\Psi=J$ in equation (\ref{cano2}) and

\begin{equation}
\Psi^\dag\eta\Psi=S=
\begin{pmatrix}
\begin{smallmatrix}
 S_{2n_1}&&&&&\\
&\ddots&&&\\
&& S_{2n_p}&&&&\\
&&&\epsilon_{n_q} S_{n_q}&&\\
&&&&\ddots&\\
&&&&&\epsilon_{n_r} S_{n_r}
\end{smallmatrix}
\end{pmatrix}\label{cano3},
\end{equation}
where $n_i$ are the orders of the Jordan blocks in equation (\ref{cano2}), $S_{k}=\begin{pmatrix}\begin{smallmatrix}&&1\\&\iddots&\\1&&\end{smallmatrix}\end{pmatrix}_{k\times k}$ and $\epsilon_{n_i}=\pm 1$ are uniquely determined by $\eta$.
\end{lem}

Lemma \ref{lem2} actually gives a standard way to construct metric operators.
Note that $S_k$ is positive definite only if $k=1$.
Utilizing this, one can verify that $\eta$ is positive definite if and only if $H$ is unbroken. Hence when $H$ is broken, the $\eta$-inner product of a state can be negative.

\subsection{An example in $\mathbb C^2$}
In $\mathbb C^2$, Lemma \ref{cano2} and Lemma \ref{cano3} will give relatively simple properties of a $\cal PT$-symmetry quantum system.

By Lemma \ref{cano2}, there are three cases of the canonical form $J=\Psi^{-1}H\Psi$ and $PT\overline{\Psi}$.

(i) $J$ is diagonal and has two real eigenvalues $\lambda_1=a_1$ and $\lambda_2=a_2$.
Let $\ket{\psi_1}$ and $\ket{\psi_2}$ be the two column vectors of $\Psi$, then we have,

\begin{equation}
\begin{split}
&H\ket{\psi_i}=a_i\ket{\psi_i},\\
&{\cal PT}\ket{\psi_i}=\ket{\psi_i} ( PT\overline{\ket{\psi_i}}=\ket{\psi_i}).
\end{split}
\label{4}
\end{equation}

(ii) $J$ has two equal real eigenvalues $\lambda_1=\lambda_2=a$ and is not diagonal.
\begin{equation}
\begin{split}
&H\ket{\psi_1}=a\ket{\psi_i},\\
&H\ket{\psi_2}=a\ket{\psi_2}+\ket{\psi_1},\\
&{\cal PT}\ket{\psi_i}=\ket{\psi_i}.
\end{split}
\label{5}
\end{equation}

(iii) $J$ is diagonal and has two complex conjugate eigenvalues $\lambda_1=\overline{\lambda_2}=a+ib$.
\begin{equation}
\begin{split}
&H\ket{\psi_i}=\lambda_i\ket{\psi_i},\\
&{\cal PT}\ket{\psi_1}=\ket{\psi_2},\\
&{\cal PT}\ket{\psi_2}=\ket{\psi_1}.
\end{split}
\label{6}
\end{equation}

One can use Lemma \ref{lem2} to construct corresponding metric operators,

(i) $\eta=(\Psi^{-1})^\dag\Psi^{-1}$.  Then we have
\begin{equation}
\braket{\psi_i|\psi_j}_\eta=\delta_{ij}.\label{7}
\end{equation}

(ii) $\eta=(\Psi^{-1})^\dag S_2\Psi^{-1}$,
\begin{equation}
\begin{split}
\braket{\psi_1|\psi_2}_\eta=\braket{\psi_2|\psi_1}_\eta=1,\\
\braket{\psi_1|\psi_1}_\eta=\braket{\psi_2|\psi_2}_\eta=0.
\end{split}
\label{8}
\end{equation}

(iii) $\eta=(\Psi^{-1})^\dag S_2\Psi^{-1}$, the $\eta$-inner product of $\psi_i$ is the same as that in (\ref{8}).

Similar to the role of standard inner product in conventional quantum physics,
 the $\eta$-inner product determine the mechanism of a quantum system in $\cal PT$-symmetry quantum mechanics.
 In conventional quantum physics, the inner product of a state is interpreted as the probability.
However, the $\eta$-inner product of a state can be negative, which obscures the physical interpretation of it.
In spite of this, an understanding from the perspective of superposition and coherence may be a candidate of such an interpretation.

\section{Metric operator and its relation with superposition and coherence}
In this part, we will show that the $\eta$-inner
 product depicts some superposition property for broken $\cal PT$-symmetric systems or some superposition-free property for unbroken
 $\cal PT$-symmetric systems.
To illustrate this, it is sufficient to discuss the $\cal PT$-symmetric systems in $\mathbb C^2$.
\subsection{The broken $\cal PT$-symmetry}

%As was mentioned, $\eta$-inner product is an important notion in $\cal PT$-symmetry quantum theory.
% In fact, utilizing the analogous viewpoints in the resource theory, one can find that the $\eta$-inner product is tightly related to
% ``superposition''. A good manifestation is the two dimensional case. We will show that the $\eta$-inner
% product depict some superposition property for broken $\cal PT$-symmetric systems while some superposition-free property for unbroken
% $\cal PT$-symmetric systems.

Let $H$ be a broken $\cal PT$-symmetric Hamiltonian and $\rho=\sum\rho_{ij}\ket{\psi_i}\bra{\psi_j}$ be a state.

If $H$ has two complex eigenvalues $\lambda$ and $\overline{\lambda}$, it follows from (\ref{6}) that

$U(t)\ket{\psi_1}=e^{-it\lambda}\ket{\psi_1},$

$U(t)\ket{\psi_2}=e^{-it\overline{\lambda}}\ket{\psi_2}.$

Let $\rho(t)=U(t)\rho U^\dag(t)=\sum\rho_{ij}(t)\ket{\psi_i}\bra{\psi_j}$.
Direct calculations show that $\rho_{12}(t)=\rho_{12}$, $\rho_{21}(t)=\rho_{21}$.
Since the superposition only concerns the coefficients of  $\ket{\psi_i}\bra{\psi_j}(i\neq j)$,
the superposition can be considered to be stationary during the evolution.

If $H$ cannot be diagonalized, it follows from (\ref{5}) that

$H=\Psi\bpm a&1\\0&a \epm\Psi^{-1}= a I_2+ \Psi\bpm 0&1\\0&0 \epm\Psi^{-1}$,

$U(t)=e^{-itH}=e^{-ita}(I-it\Psi\bpm 0&1\\0&0 \epm\Psi^{-1})$,

$e^{-itH}\ket{\psi_1}=e^{-ita}\ket{\psi_1}$,

$e^{-itH}\ket{\psi_2}=e^{-ita}\ket{\psi_2}-ite^{-ita}\ket{\psi_1}$.

Direct calculations show that $\rho_{12}(t)\neq \rho_{12}$ and $\rho_{21}(t)\neq\rho_{21}$.
However, $\rho_{12}(t)+\rho_{21}(t)=\rho_{12}+\rho_{21}$. Although the superposition is not stationary, the ``sum of
 superposition'' is preserved.

That is, in some sense, the
superposition of a state are preserved by the action of $\cal PT$-symmetric evolution.
Furthermore, let $\ket{\xi}=b_1\ket{\psi_1}+b_2\ket{\psi_2}$ be a state.
(\ref{8}) shows that $\braket{\xi,\xi}_{\eta}=b_1\overline{b_2}+b_2\overline{b_1}=\rho_{12}+\rho_{21}$.
Note that $\braket{\xi,\xi}_{\eta}=\braket{\xi,\eta \xi}=Tr(\eta \ket{\xi}\bra{\xi})$. Similarly, for a general state $\rho$,
$Tr(\eta\rho)=\rho_{12}+\rho_{21}$.
 Hence, $\eta$-inner product is actually the
`` the sum of superposition''.
In addition, for two states $\ket{\xi_1}$ and $\ket{\xi_2}$, $\braket{\xi_1,\xi_2}_{\eta}$ is the ``sum of superposition'' of
$\ket{\xi_2}\bra{\xi_1}$.

Thus, the $\eta$-inner product is not the probability in the usual sense but some description of the superposition.
This also gives another way to understand why the $\eta$-inner product can be negative for broken $\cal PT$-symmetric systems.
Although the positivity of probability is natural, the positivity of a quantity concerning
superposition is unnatural and not necessary, since superposition is a phenomenon in quantum theory,
 which does not have a sign itself.

 Intuitively, a process conserving the superposition for any state may not exist in the framework of conventional quantum mechanics,
inferring the impossibility of simulating a broken $\cal PT$-symmetric Hamiltonian in the usual sense.
In fact, it can be rigorously showed that one cannot simulate a broken $\cal PT$-symmetric Hamiltonian by utilizing a large Hermitian Hamiltonian
\cite{huang2017embedding}.

\subsection{The unbroken $\cal PT$-symmetry}
When $H$ is unbroken,
$\rho(t)=U(t)\rho U^\dag(t)=\sum \rho_{ij}e^{it(\lambda_j-\lambda_i)}\ket{\psi_i}\bra{\psi_j}$.
Since $\lambda_1\neq\lambda_2$ in general, it is apparent that $\rho_{12}(t)\neq \rho_{12}$, $\rho_{21}(t)\neq\rho_{21}$ and $\rho_{12}(t)+\rho_{21}(t)\neq\rho_{12}+\rho_{21}$.
However, in this case, $\rho_{11}(t)=\rho_{11}$ and $\rho_{22}(t)=\rho_{22}$.
Moreover, $Tr(\eta \rho)=\rho_{11}+\rho_{22}=Tr(\eta \rho(t))=\rho_{11}(t)+\rho_{22}(t)$.
In particular, for a pure state $\ket{\xi}=b_1\ket{\psi_1}+b_2\ket{\psi_2}$,
then $\braket{\xi,\xi}_{\eta}=Tr(\eta\ket{\xi}\bra{\xi})=|b_1|^2+|b_2|^2=\rho_{11}+\rho_{22}$.
Compared with the broken case, $\eta$-inner product now characterizes the superposition-free properties of a state.

It is also convenient to discuss the effect of the metric operator $\eta$
when $H$ is unbroken. Note that $\psi_1$ and $\psi_2$ are not orthogonal in the
standard inner product. However, $\braket{\psi_i|\psi_j}_\eta=\delta_{ij}$, which shows the orthogonality. Thus, the metric operator $\eta$ mathematically transforms the superposition to coherence.
If we do not consider the normalization, a superposition-free state $\rho=\sum p_i\ket{\psi_i}\bra{\psi_i}$ in the usual sense
is incoherent with respect to the $\eta$-inner product.
Such a change of inner product is not unitary, if it can be realized in the framework of conventional quantum mechanics, the process
can only be probabilistically.

In fact, it is possible to simulate
the transformation $\rho\mapsto U(t)\rho U^\dag(t)$ in a subsystem of a large Hermitian system.
To see this, note that $U(t)=e^{-itH}=\Psi e^{-it\Lambda}\Psi^{-1}$, where $\Lambda=\bpm\begin{smallmatrix}\lambda_1&0\\0&\lambda_2\end{smallmatrix}\epm$.
Thus $\|U(t)\|=\|\Psi e^{-it\Lambda}\Psi^{-1}\|\leqslant \|\Psi\|\|\Psi^{-1}\|$. Hence $U(t)$ forms a set of uniformly bounded operators
and it is possible to find some constant $c$ such that $c^2U^\dag(t)U(t)\leqslant I$. In fact, using the Naimark Dilation method one can always
realize a transformation $\rho\mapsto \frac{U(t)\rho U^\dag(t)}{Tr[U(t)\rho U^\dag(t)]}$. For concrete discussions, see
\cite{gunther2008naimark,huang2017embedding}.

Furthermore, note that $cU(t)=c\Psi e^{-it\Lambda}\Psi^{-1}=\sum ce^{-it\lambda_i}\ket{\psi_i}\bra{\psi^\perp_i}$.
Define the free states to be $\rho=\sum p_i\ket{\psi_i}\bra{\psi_i}$.
It is apparent that for any free state $\rho$, the state $\frac{U(t)\rho U^\dag(t)}{Tr[U(t)\rho U^\dag(t)]}$ is also free.
Hence $cU(t)$ will gives a free operation.
Such a result is no coincidence. As was mentioned, $\eta$-inner product is an invariant superposition-free quantity, which reflects the
 stationary superposition-free property of the Hamiltonian $\cal H$. The free property of $cU(t)$ is just another manifestation.
This conclusion is also true in any other finite dimensional space.

The different superposition properties of $\eta$-inner product and $\cal PT$-symmetric Hamiltonian $\cal H$
itself, essentially arises from the $\cal PT$-symmetry.
Intuitively, for an anti-linear operator $\cal PT$ and the basis vectors $\{\ket{\psi_i}\}$, the condition ${\cal PT}\ket{\psi_i}=\ket{\psi_i}$ in (\ref{4}) reflects the interrelation of a basis vector with itself under the action of $\cal PT$.
Through the condition ${[\cal H, PT]}=0$,
such interrelations are respected and manifested by the property of unbroken $\cal PT$-symmetric system.
However, ${\cal PT}\ket{\psi_1}=\ket{\psi_2}$, ${\cal PT}\ket{\psi_2}=\ket{\psi_1}$ in (\ref{6}) is also possible.
 This can be viewed as
the interrelation of a basis vector with other basis vector under the action of $\cal PT$, just like the coherence or superposition.
Similarly, the interrelations are also preserved in the broken $\cal PT$-symmetric system, in a form of superposition.
(\ref{5}) actually gives a intermediate case between (\ref{4}) and (\ref{6}), in which
under the action of $H$, $\ket{\psi_1}$ interrelates with itself while $\ket{\psi_2}$ interrelates with both $\ket{\psi_i}$.
A support of this viewpoint is Bender's model, in which $H=\bpm re^{i\theta}&s\\s&re^{-i\theta}\epm$ is the $\cal PT$-symmetric Hamiltonian,
 where $r$, $s$, $\theta$ are real parameters \cite{bender2007making}. When $s^2-r^2\sin^2\theta>0$, $H$ is unbroken, corresponding to (\ref{4}). When
$s^2-r^2\sin^2\theta<0$, $H$ has two complex conjugate eigenvalues, corresponding to (\ref{6}). The intermediate condition
 $s^2-r^2\sin^2\theta=0$ correspond to a case in which $H$ cannot be diagonalized in general, namely, the case of (\ref{5}).

\section{Discussions}
In this section, we aim to investigate our results from the perspective of optics. We will see a tight connection between the Stokes
parameters and the $\eta$-inner product.

In $\mathbb C^2$, one can easily map the states as the two degree of freedoms in optical polarizations, i.e., $|\psi_i\rangle$ being the polarization in the vertical or horizontal direction, denoted as $x$ and $y$ coordinates.
For broken $\cal PT$-symmetry, the $\eta$-inner product gives the quantity $b_1 \overline{b_2} + b_2\overline{b_1}$, which is nothing but the {\it Stokes parameters}, $S_2 = E_x \overline{E_y} + E_y \overline{E_x}$~\cite{Stokes}.
Here, the polarization components in the $x$ and  $y$ directions are denoted by $E_x$ and $E_y$, respectively.
It is known that Stokes parameter, $S_2$ measures the degree of polarization.
However, for unbroken $\cal PT$-symmetry, the  $\eta$-inner product gives the quantity $\rho_{11} + \rho_{22}$, which is an analogue to another Stokes parameter, $S_0 = |E_x|^2 + |E_y|^2$, corresponding to the total intensity (here the probability) of the field.

Nevertheless, for the non-diagonalizable case, there is only one eigenstate, failing to fulfill the Stokes parameterization of its own.
In this case, one more basis vector is needed, which corresponds to the non-diagonalizable Hamiltonian.
Even though the Stokes parameters $S_0$ and $S_2$ are essentially different, for such a non-diagonalizable case, one can transfer in different Stokes parameters,  from $S_0$ to $S_2$, or vice versa.

As an example, consider the $\cal PT$-symmetric Hamiltonian $H=\begin{pmatrix} re^{i\theta}&s\\s&re^{-i\theta}\end{pmatrix}$.
The two eigenstates of $H$ are $\ket{\tilde{E}_+(\alpha)}=\frac{1}{\sqrt{2}}\bpm e^{\frac{i\alpha}{2}}\\e^{-\frac{i\alpha}{2}}\epm$ and
$\ket{\tilde{E}_-(\alpha)}=\frac{1}{\sqrt{2}}\bpm ie^{-\frac{i\alpha}{2}}\\-ie^{\frac{i\alpha}{2}}\epm$, where $\sin\alpha=\frac{r}{s}\sin\theta$.
Since $\braket{\tilde{E}_\pm(\alpha)|\tilde{E}_\pm(\alpha)}_\eta=\cos\theta$, consider the two $\eta$-inner product normalised states,
$\ket{E_\pm(\alpha)}=\frac{1}{\sqrt{\cos\alpha}}\ket{\tilde{E}_\pm(\alpha)}$.
For a state $\ket{\xi}=\bpm x\\y \epm=c_1\ket{E_+(\alpha)}+c_2\ket{E_-(\alpha)}$, we can obtain the coefficients $c_i$,
\begin{eqnarray*}
&&c_1=\sqrt{2\cos\alpha}\frac{xe^{i\frac{\alpha}{2}}+ye^{-i\frac{\alpha}{2}}}{e^{i\alpha}+e^{-i\alpha}},\\
&&c_2=-i\sqrt{2\cos\alpha}\frac{xe^{-i\frac{\alpha}{2}}-ye^{i\frac{\alpha}{2}}}{e^{i\alpha}+e^{-i\alpha}},
\end{eqnarray*}
Moreover,
\begin{eqnarray}
\nonumber S_0&=&|c_1|^2+|c_2|^2\\
&=&\frac{1}{\cos\alpha}(|x|^2+|y|^2+i(x\bar{y}-y\bar{x})\sin\alpha).
\end{eqnarray}
Note that the $\cal PT$-symmetry breaking condition is $s^2-r^2\sin^2\theta=0$. Hence $\alpha=\frac{\pi}{2}$ is a critical point.
As $\alpha\rightarrow \frac{\pi}{2}$, $S_0=|c_1|^2+|c_2|^2\rightarrow \infty$. To see what actually happens at the critical point of $H$, note that the two eigenstates $\ket{\tilde{E}_\pm(\frac{\pi}{2})}$ coincide.
So the eigenstate only gives one direction, failing to fully realize the Stokes parameterization. This calls for one more basis vector, which is the generalized eigenstate given by (\ref{5}), leading to a non-trivial non-diagonalizable case. In addition, utilizing the new state and such a non-diagonalizable Hamiltonian makes us transfer from $S_0$ to $S_2$, which are essentially different Stokes parameters. And the discussion of $S_2$ in the broken case is an analogy to the above.

Nevertheless, when we generalize the results to $\mathbb C^n$, a simple optical interpretation becomes unclear.
As the scenario for Stokes parameters, the optical interpretation for our $\eta$-inner product shares the same problem of complications  in higher dimensions.

\section{Conclusion}
In this note, we discuss $\cal PT$-symmetry theory in the view of superposition and coherence.
A more physical interpretation of $\eta$-inner
product is given. This shows the physical difference between the broken and unbroken $\cal PT$-symmetric systems. We also argue that the
essence of such a interpretation comes from the $\cal PT$-symmetry of a system, which is natural according to our
physical intuitions. Though the discussions are restricted to $\mathbb C^2$, it is possible to generalize the idea by utilizing Lemma \ref{lem1} and Lemma \ref{lem2}. In that case, for a state $\rho=\sum \rho_{ij}\ket{\psi_i}\bra{\psi_j}$, only part of the $\rho_{ij}(i\neq j)$ will be involved. However, the interpretation of $\eta$-inner
product as stationary superposition or superposition-free property is still valid.

\begin{acknowledgments}
The project is supported by National Natural Science Foundation of China (11171301, 11571307).
 \end{acknowledgments}

%\bibliographystyle{apsrev4-1}
%\bibliography{index}

%

\end{document}